\documentclass[conference,compsoc]{IEEEtran}

\usepackage{amsmath,amssymb,amsfonts}
\usepackage{algorithmic}
\usepackage{graphicx}
\usepackage{textcomp}
\usepackage{xcolor}

\usepackage{float}
\usepackage[english]{babel}
\usepackage[sorting=none,style=numeric]{biblatex}

\begin{document}

\title{A Review on the State of the Art in Non Contact Sensing for COVID-19}

\author{\IEEEauthorblockN{William Taylor}
\IEEEauthorblockA{James Watt School of \\ Engineering,\\ University of Glasgow, \\Glasgow, United Kingdom}

\and

\IEEEauthorblockN{Qammer H. Abbasi}
\IEEEauthorblockA{James Watt School of \\ Engineering,\\ University of Glasgow, \\Glasgow, United Kingdom}

\and

\IEEEauthorblockN{Kia Dashtipour}
\IEEEauthorblockA{James Watt School of \\ Engineering,\\ University of Glasgow, \\Glasgow, United Kingdom}

\and

\IEEEauthorblockN{Shuja Ansari}
\IEEEauthorblockA{James Watt School of \\ Engineering,\\ University of Glasgow, \\Glasgow, United Kingdom}

\and

\IEEEauthorblockN{Aziz Shah}
\IEEEauthorblockA{James Watt School of \\ Engineering,\\ University of Glasgow, \\Glasgow, United Kingdom}

\and

\IEEEauthorblockN{Arslan Khan}
\IEEEauthorblockA{James Watt School of \\ Engineering,\\ University of Glasgow, \\Glasgow, United Kingdom}

\and

\IEEEauthorblockN{Muhammad Ali Imran}
\IEEEauthorblockA{James Watt School of \\ Engineering,\\ University of Glasgow, \\Glasgow, United Kingdom}

}

\maketitle

\begin{abstract}
COVID-19 disease, caused by SARS-CoV-2, has resulted in a global pandemic recently. With no approved vaccination or treatment, governments around the world have issued guidance to their citizens to remain at home in efforts to control the spread of the disease. The goal of controlling the spread of the virus is to prevent strain on hospital. In this paper, we have focus on how non-invasive methods are being used to detect the COVID-19 and assist healthcare workers in caring for COVID-19 patients. Early detection of the COVID-19 virus can allow for early isolation to prevent further spread. This study outlines the advantages and disadvantages and a breakdown of the methods applied in the current state-of-the-art approaches. In addition, the paper highlights some future research directions, which are required to be explored further to come up with innovative technologies to control this pandemic.
\end{abstract}

\section{Introduction}
Since late 2019, countries around the world have been experiencing a global pandemic through the surfacing and spread of the potentially fatal COVID-19 (COronaVIrusDisease 2019) caused by SARS-CoV-2 (Severe Acute Respiratory syndrome CoronaVirus 2) virus \cite{singhal2020review}. The COVID-19 causes victims to develop a fever and display respiratory difficulties causing coughing or shortness of breath \cite{pan2020clinical, poyiadji2020covid, xu2020pathological}. Data collected from victims of the virus shows that the majority of deaths occurred in patients with underlying health issues with elderly people being at a higher risk of death \cite{wu2020exposure}. The first confirmed case of the virus is considered to be in Wuhan, China in December 2019 with some of the early cases thought to be traced to seafood markets trading live animal species such as bats and snakes \cite{hellewell2020feasibility,jiang2020novel,khan2020modeling,nishiura2020initial}. The virus has been discovered to likely be related to bats. It is suspected that the virus may have transmitted to humans through bats which were being sold as food items \cite{shereen2020covid, lai2020severe}. The exact cause of the virus is still unknown and it has also been suggested that the virus could originate from pangolins, which are natural hosts of corona viruses \cite{liu2020pangolins}. Pangolins is unlikely linked to the outbreak as the corona viruses found on the animal differ to COVID-19 \cite{xiao2020isolation}. Although it is possible the pangolin could have served as a intermediate host. As a result, these markets were shutdown in China \cite{novel2020epidemiological}. The virus rapidly spread throughout China and eventually spread throughout the world. The virus was officially declared a global pandemic by the world health organisation (WHO) on 30th January 2020 \cite{spinelli2020covid, zheng2020covid}. Although new discoveries are being made at the time of writing this paper, the virus has been found to be highly contagious and this has led to the rapid spread throughout the world \cite{dong2020role}. The virus is spread primarily through respiratory droplets from an infected person \cite{cai2020indirect}. These droplets can be dispensed by an infected person when coughing or sneezing. The droplets can then infect others directly via the eyes, mouth or nose when they are within a one meter radius of an infected person. The droplets can also be passed to others indirectly due to their long-term presence on surfaces \cite{world2020modes}. Another leading factor in the rapid spread is that those infected with COVID-19, can be contagious during the early stages of infection while they are showing no symptoms \cite{kooraki2020coronavirus}. This leads to people believing they are not sick while unknowingly spreading the virus. One of the main challenges of the COVID-19 pandemic is the how the spread of the virus can be controlled. The rapid spread of the COVID-19 virus has highlighted how the world's population interacts when faced with a pandemic \cite{dowd2020demographic}. Governments around the world have outlined guidelines to their citizens to adhere to lock down rules. Currently, the best strategy to control the spread of COVID-19 is to ensure socially distancing people until a vaccine or an effective treatment can be produced \cite{salathe2020covid,lewnard2020scientific}. The National Health Service (NHS) of the United Kingdom is expecting an increased demand for their services as more COVID-19 patients are admitted and staff sick leave increases as staff members contract the disease \cite{willan2020challenges}. Technology is being rapidly introduced in healthcare applications to develop systems that can ease the demand of the health service \cite{yang2020human,abbasi2016advances, taylor2020intelligent}. Any assistance via healthcare technology will free up valuable clinical resources to focus on other areas of care. In this paper we look at the state of the art non-contact sensing techniques and how these technologies can be used to assist in the care and detection of people suffering from COVID-19. 

\section{Non-contact sensing to detect COVID-19 symptoms}
Non contact sensing is the ability to detect information without direct contact with a subject. In terms of healthcare, non contact can be used monitor the human body without devices physically touching the body. Non contact techniques are considered highly valuable in dealing with a highly infectious disease such as COVID-19 as contact may contribute to the spread of disease. Healthcare sensing technologies aim to collect information from a person which can be processed by Artificial Intelligence (AI) to provide decision support or directly analysed by a clinician to diagnose a disease or monitor existing conditions. Non-contact remote sensing technology is able to sense such healthcare markers without introducing anything to the body (e.g. wearable devices). Wearable devices can be uncomfortable for some which will entice users to remove the device and results in misplacement or damage \cite{tan2018exploiting}. The non-contact techniques can assist in the detection of COVID-19 and the care of patients suffering from COVID-19. Vital sign monitoring can provide great assistance in the fight against COVID-19 for a number of reasons. Although COVID-19 affects the respiratory system \cite{marini2020management,fan2018breathing}, it has also been shown to take affect on the cardiovascular system \cite{zheng2020covid}. Therefore non-contact sensing that monitors these vital signs can be used to aid in the detection and treatment of COVID-19. Examples of Non-contact techniques described in this paper include computed tomography (CT) scans, X-rays, Camera Technology, Ultrasound Technology, Radar Technology and Radio Frequency (RF) signal sensing. Table \ref{Techniques} details the advantages and disadvantages of each technique. Table \ref{literature} provides a summary table of the current literature contained within this review paper.

\begin{table}[htb]
\caption{Summary of Non-Invasive Techniques} \label{Techniques}
\centering
\begin{tabular}{||p{1cm} p{3cm} p{3cm}||}  
 \hline
Non-Invasive Techniques & Advantages & Disadvantages \\ [0.5ex] 
 \hline\hline
 CT Scans & High Accuracy,  & High cost,\\
          & High image resolution & Exposure to Radiation,\\ 
          &                       & Non-portable,  \\
          &                       & Professional required for image analysis\\
 \hline
 X-rays & High Accuracy,  &  High cost,\\
        & High image resolution & Non-portable,\\
        &                       & Exposure to Radiation, \\
                                &&Professional required for image analysis\\
 \hline
 Camera & High Accuracy, & High cost of Camera\\
 Technology                    & AI can be used for data analysis & Camera needs to be operated by professional\\
 \hline
 Ultrasound & High Accuracy, & Patients need to\\
 Technology & AI can be used for data analysis & be prepared before \\
            & Possibility of portability & Ultrasound scan\\
 \hline
 Radar & High accuracy,  & High power consumption, \\
  Technology               & AI can be used for data analysis & High cost\\ 
 \hline
  RF signals  & Low cost, & Vulnerable to noise\\
  &High Accuracy &\\ 
  &AI can be used for data analysis &\\ 
 \hline
 \end{tabular}
\end{table}

\begin{table*}[htb]
\caption{Summary of Current Literature} \label{literature} 
\centering
\begin{tabular}{||p{4cm} p{3cm} p{5cm} p{4cm}||}   
 \hline
Title of Paper & Author(s) and Year & Key themes & Authority \\ [0.5ex] 
 \hline\hline
 Coronavirus (covid-19) classification using ct images by machine learning methods & Barstugan, M., Ozkaya, U., \& Ozturk, S. (2020), & CT scan images are used to experiment with various methods of feature extraction and deep learning algorithms to achieve the best results & Peer reviewed paper. 157 citations on Google scholar.
157 citations on Google scholar.
 \\
 \hline
 POCOVID-Net: automatic detection of COVID-19 from a new lung ultrasound imaging dataset (POCUS) & Born, J., Brändle, G., Cossio, M., Disdier, M., Goulet, J., Roulin, J., \& Wiedemann, N. (2020) & Lung Ultrasound videos of COVID-19, Pneumonia and non-infected patients used deep learning for classification. & Peer   reviewed   paper. 2 citations on Google scholar.
  \\
 \hline 
 CSAIL device lets doctors monitor COVID-19 patients from a distance   & Conner-simons 2020 & Radio Frequencies have been used to monitor the vital signs of patients in a contactless manner to protect healthcare workers & Article found on MIT Computer Science \& Artificial Intelligence Laboratory website.
 \\
 \hline 
Can Radar Remote Life Sensing Technology Help to Combat COVID-19?     & Islamcan 2020 & Radar systems have been used to monitor the vital signs of patients in a contact less manner to protect healthcare workers & Paper uploaded on researchgate.net.
 \\
 \hline 
 Combining  Visible  Light  and  Infrared  Imaging  for  Efficient  Detection  of  Respiratory Infections such as COVID-19 on Portable Device & Jiang, Z., Hu, M., Fan, L., Pan, Y., Tang, W., Zhai, G., \& Lu, Y. (2020) & RGB-Terminal camera footage used in a BiGRU neural network model between healthy and ill. & Peer reviewed paper.
  \\
 \hline 
 Automatic detection of coronavirus disease (covid-19) using x-ray images and deep convolutional neural networks & Narin, A., Kaya, C., \& Pamuk, Z. (2020) & X-ray scan images are used in a ResNet-50 Convolutional Neural Network (CNN)  to distinguish between COVID-19 and Non-infected scan images. & Peer reviewed paper. 102 citations on Google scholar.
 \\
 \hline 
  Artificial intelligence distinguishes COVID-19 from community acquired pneumonia on chest CT    & Li, L., Qin, L., Xu, Z., Yin, Y., Wang, X., Kong, B., \& Cao, K. (2020) & CT scan images are used in a COVNet neural network to distinguish between COVID-19, Pneumonia and Non-infected scan images. & Peer reviewed paper. 157 citations on Google scholar.
 \\
 \hline 
 Automated detection of COVID-19 cases using deep neural networks with X-ray images    & Ozturk, T., Talo, M., Yildirim, E. A., Baloglu, U. B., Yildirim, O., \& Acharya, U. R. (2020) & Xray images are processed using the DarkNet neural network to test binary classification between COVID and Non-infected and multiclass classification between COVID, Pneumonia and Non-infected. & Peer reviewed paper. 22 citations on Google scholar.
 \\
 \hline 
 Lung infection quantification of COVID-19 in ct images with deep learning    & Shan, F., Gao, Y., Wang, J., Shi, W., Shi, N., Han, M., ... \& Shi, Y. (2020) & CT scan images are used in Deep learning to identify COVID-19. Human in the loop technique is used to focus on increasing accuracy & Peer reviewed paper. 52 citations on Google scholar.
\\
 \hline 
 Abnormal respiratory patterns classifier may contribute to large-scale screening of people infected with COVID-19 in an accurate and unobtrusive manner     & Wang, Y., Hu, M., Li, Q., Zhang, X. P., Zhai, G., \& Yao, N. (2020) & The paper details that COVID-19 patients display Tachypnea (Rapid breathing). The paper looks at taking Depth images to identify the breathing patterns of volunteers using deep learning & Peer reviewed paper. 24 citations on Google scholar.
\\
 \hline 
 Covid-19 screening on chest x-ray images using deep learning-based anomaly detection    & Zhang, J., Xie, Y., Li, Y., Shen, C., \& Xia, Y. (2020) & X-ray images are used with deep learning to identify if samples are COVID-19 or Pneumonia & Peer reviewed paper. 32 citations on Google scholar.
\\
 \hline\hline
 \end{tabular}
\end{table*}

\subsection{CT Scanning}
An example of a non invasive technique to detect COVID-19 is using Computed Tomography (CT) scans \cite{yang2020role}. This process involves taking several X-ray images of a persons chest to create a 3D image of the lungs. The images can be reviewed by professionals to look for abnormalities in the lungs. The activity in the lungs is more prominent in the later stages of infection, however ultimately the CT scans showed a sensitivity of 86 \% - 98 \% \cite{udugama2020diagnosing}. This technique is non-contact as nothing is directly introduced into the body of the patient. CT scans are able to achieve high precision with high image resolution however the technology used to perform CT scans is expensive. The equipment is not portable and it requires skilled professional for image analysis. Another disadvantage of CT scanning is that the patient is exposed to radiation \cite{ceniccola2019current}. The radiation levels in CT scans have been found to result in an estimated cancer mortality risk of 0.08\%  within a 45 year old adult \cite{brenner2004radiation}. Recently, AI has been used on CT images for diagnosis of COVID-19 \cite{shi2020review}. 

\begin{figure}[htb]
    \centering
    \includegraphics[width=0.5\textwidth]{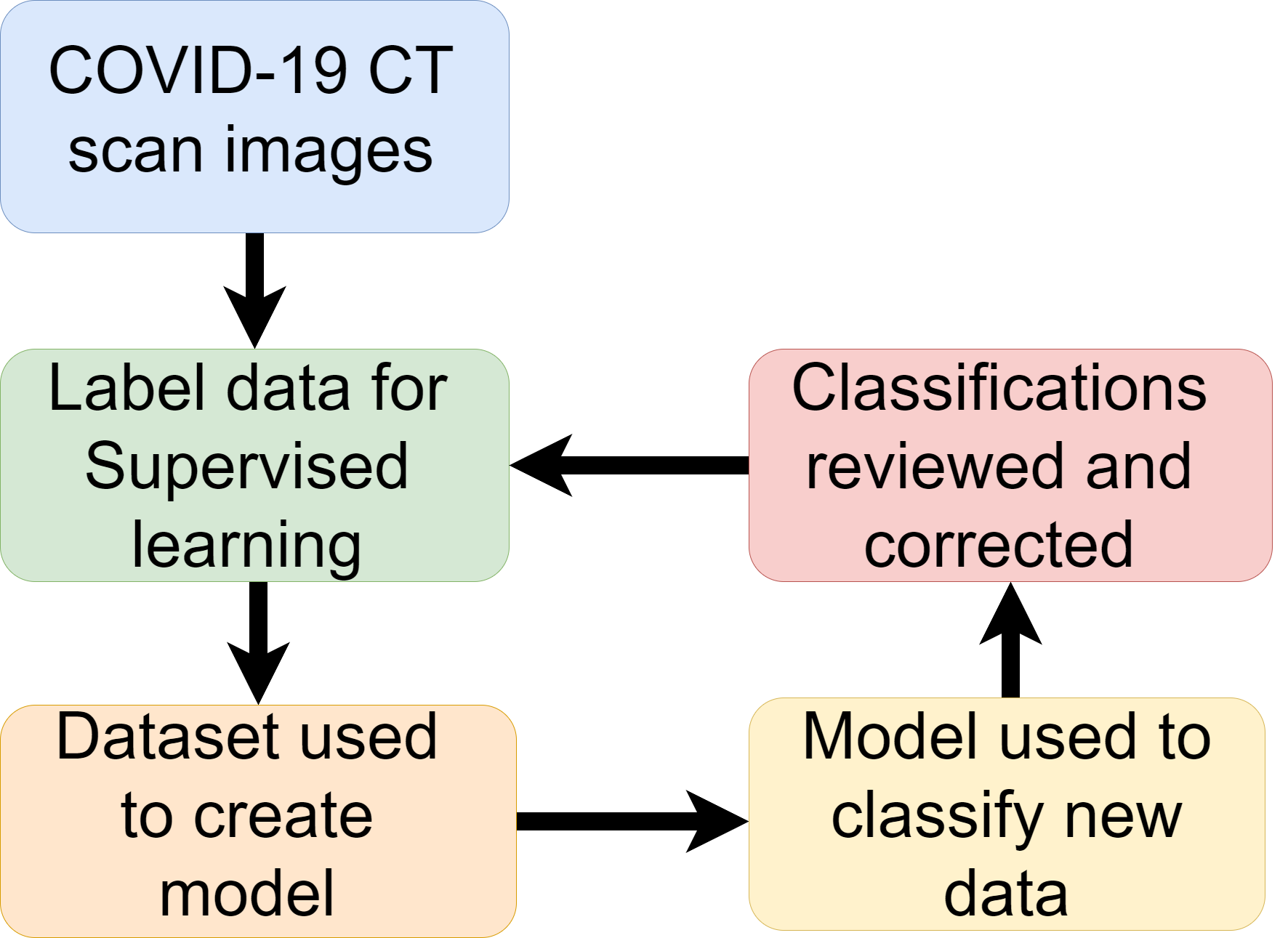}
    \caption{Flow chart of work for detection of COVID-19 from CT scan (Reproduced from \cite{shan2020lung})}
    \label{shan2020lung1}
\end{figure}

Fei Shan et al. \cite{shan2020lung} developed a deep learning model which was able to detect COVID-19 and also the level of infection. Their model adopted human-in-the-loop (HITL) strategy. HITL is when specialists are used to label a small amount of training data. Then an initial model is trained. Then this initial model is then used to classify new data. The specialist then correct any incorrect labels and the dataset can be used to train further models. This task can be iterated numerous times to reduce the tedious task of labeling large amounts of data. The experiment used 249 confirmed cases of COVID-19 for training. The experiment achieved a high result of 91.6 \% accuracy. The experiment of this paper used 3 iterations. The first iteration made classifications on the validation data using 36 labeled images as a dataset with an an accuracy score of 85.1 \%. The labels are then corrected and added to the second iteration. The second iteration used 114 images for training and achieved an accuracy result of 91.0 \%. The labels are then corrected and passed to the third iteration The third iteration is used on all 249 training images and achieved an accuracy result of 91.6 \%. The improved accuracy greatly reduces the human involvement and time devoted to labeling the full data. Figure \ref{shan2020lung1} displays a flow chart of the process of HITL.

\begin{figure}[htb]
    \centering
    \includegraphics[width=0.5\textwidth]{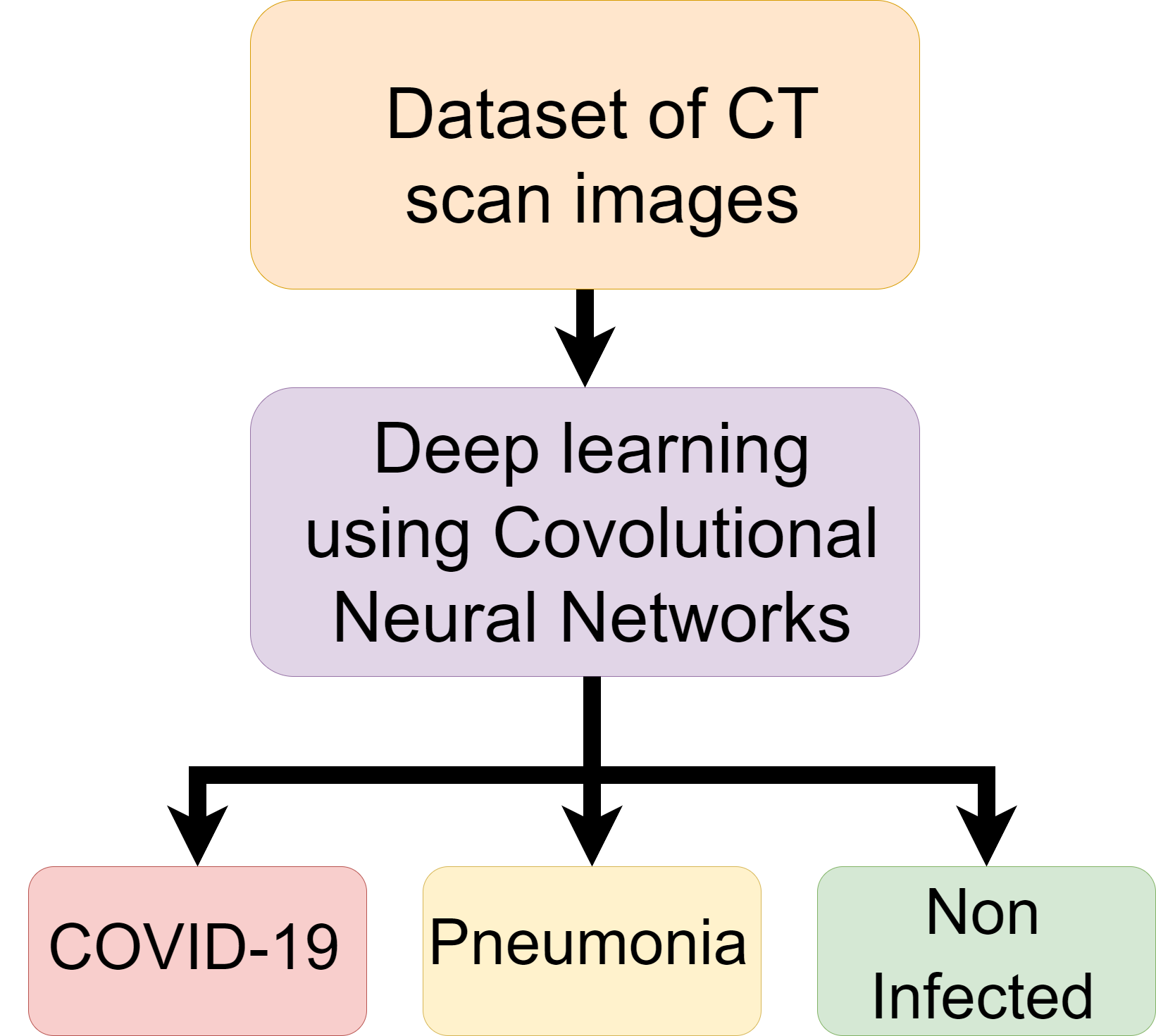}
    \caption{Flow chart of work for detection of COVID-19 from CT scan (Reproduced from \cite{li2020artificial})}
    \label{li2020artificial1}
\end{figure}

Li, Lin, et al. \cite{li2020artificial} used a COVNet - a deep learning neural network to predict COVID-19. The data used included 400 COVID-19 CT images, 1396 Pneumonia CT images and 1173 non infected CT images. The model takes CT images as input and extracts Pneumonia features from the images. The features are combined and the neural network is applied to make predictions on COVID-19, Pneumonia or non infected. Results found that the model was able to predict COVID-19 in patients with 90 \% sensitivity. The model proved to not only be able to detect infected and non infected lungs but was also able to differentiate between COVID-19 and Pneumonia \cite{li2020artificial}. Figure \ref{li2020artificial1} shows the process followed in this research. 

\begin{figure}[htb]
    \centering
    \includegraphics[width=0.5\textwidth]{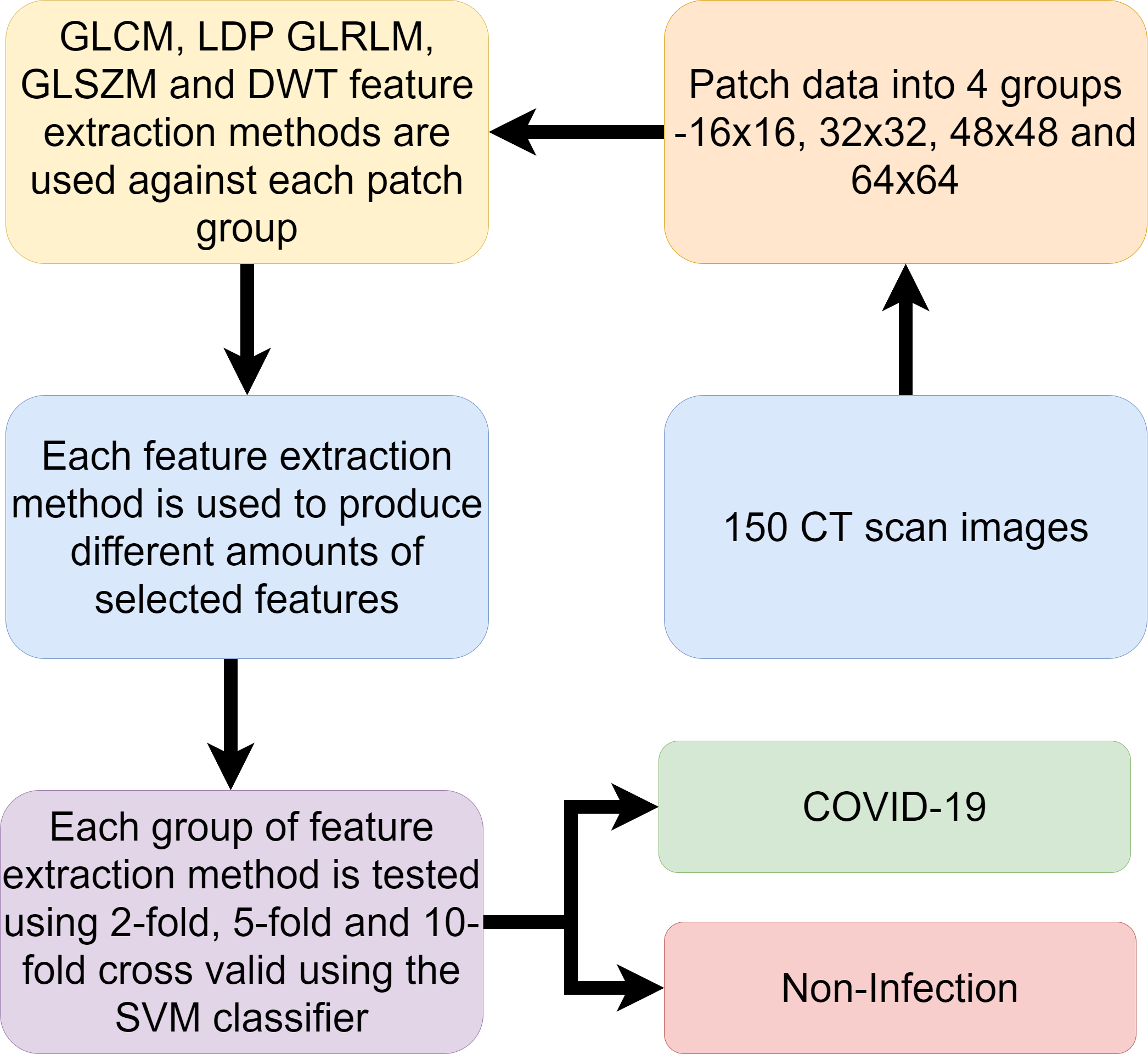}
    \caption{Flow chart of work for detection of COVID-19 from CT scan (Reproduced from \cite{li2020artificial})}
    \label{barstugan2020coronavirus1}
\end{figure}

The research of Barstugan, Mucahid et al. \cite{barstugan2020coronavirus} used machine learning on a dataset of 150 CT images. The dataset contains 53 infected cases. Patches of the images are taken. Different sized patches are used to create 4 different samples of different sized patches. The patch sizes are 16x16, 32x32, 48x48 and 64x64. The images were labelled as infected and non-infected in regards to COVID-19. The research used different methods of feature extraction on the images. These methods include Grey Level Co-occurrence Matrix (GLCM), Local Directional Patterns (LDP), Grey Level Run Length Matrix (GLRLM), Grey Level Size Zone Matrix (GLSZM) and Discrete Wavelet Transform (DWT). Support Vector Machine (SVM) algorithm was then used to classify the extracted features of each of the methods. SVM was experimented on the features using 2-fold, 5-fold and 10-fold cross validation. The highest accuracy result achieved was 99.64 \%. This result was achieved using DWT with 10 fold cross validation using 48x48 patch dimension images. A flow chart of the methodology followed in this research is shown in figure \ref{barstugan2020coronavirus1}.

The above papers show that CT scanning is capable of showing the signs of COVID-19 within a persons lungs. The research has also shown how AI can be used to assist in the determination of whether COVID-19 is present in the lungs. The studies have also shown how AI can also determine the level of infection and differentiate between Pneumonia and COVID-19 infections. Table \ref{CTscans} provides a break down of the above research papers.

\begin{table}[htb]
\caption{Summary of CT Scanning works} \label{CTscans}
\centering
\begin{tabular}{||p{2cm} p{2.5cm} p{1.5cm} p{2cm}||}  
 \hline
Title of Paper & Training data & Algorithms & Results \\ [0.5ex] 
 \hline\hline
Lung infection quantification of covid-19 in ct images with deep learning & 249 CT images of COVID-19 showing different levels of infection.& Custom Convolutional neural network (CNN) called “VB-Net”& 91.6 \% Accuracy\\ 
 \hline
Artificial intelligence distinguishes COVID-19 from community acquired pneumonia on chest CT&400 COVID-19 CT images, 1396 Pneumonia CT images and 1173 non infected CT images & Custom Convolutional neural network (CNN) called “COVNet”&90 \% sensitivity of COVID-19 samples. \\ 
 \hline
Coronavirus (covid-19) classification using ct images by machine learning methods & 150 CT images including 53 COVID-19 cases. &Support Vector Machine  & 99.64 \% Accuracy \\ 
 \hline\hline
 \end{tabular}
\end{table}

\subsection{X-ray Imaging}
X-ray images are able to provide an analysis of the health of the lungs and is used frequently to diagnose pneumonia \cite{sethy2020detection}. The same strategy is used with X-ray images of the lungs to display the visual indicators of COVID-19 \cite{wang2020covid, ghoshal2020estimating}. Similar to CT scans, X-ray equipment is also expensive and requires professionals to analyse the X-ray image.

The paper entitled "Automatic detection of coronavirus disease (covid-19) using x-ray images and deep convolutional neural networks" used X-ray images of COVID-19 infected and non infected patients to create a dataset which was used to predict COVID-19 automatically in patients. The X-ray images are used in a ResNet-50 Convolutional Neural Network (CNN) which successfully obtained results of 98 \% accuracy \cite{narin2020automatic}. 

The paper of Zhang, Jianpeng, et al. \cite{zhang2020covid} used Deep learning on a dataset of X-ray images of 70 patients confirmed to have COVID-19. Additional images of patients with Pneumonia are added from a public chest X-ray dataset. The model is used to identify between patients with COVID-19 and Pneumonia. The proposed deep learning model was able to achieve a sensitivity of 90 \% detecting COVID-19 and a specificity of 87.84 \% in detecting non COVID-19 cases. 

\begin{figure}[htb]
    \centering
    \includegraphics[width=0.5\textwidth]{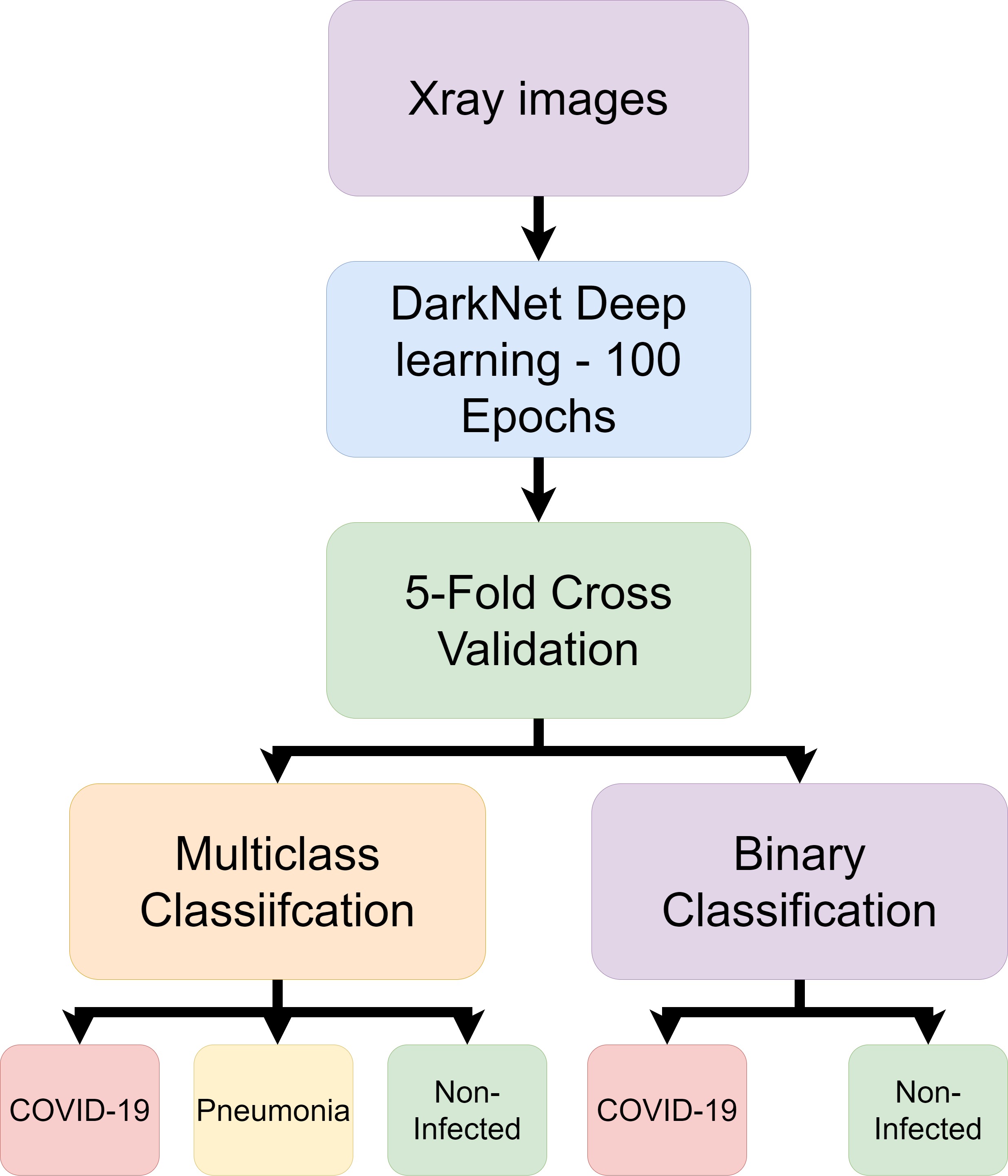}
    \caption{Flow chart of work for detection of COVID-19 from X-ray Image (Reproduced from \cite{ozturk2020automated})}
    \label{ozturk2020automated1}
\end{figure}

Ozturk, Tulin, et al. \cite{ozturk2020automated} also conducted experiments of using deep learning to classify X-ray images of patients. The experiments made use of a custom deep learning model named DarkNet to perform binary and multi-class classifications. The binary classification seeks to distinguish between COVID-19 and no findings of disease. The multi-class classification distinguishes between no findings of disease and the presence of Pneumonia. The experiments used a publicly available dataset of COVID-19 X-ray scans and another publicly available dataset for non-infected and pneumonia X-ray scans. The complete dataset included 127 COVID-19 samples and 500 each of non-infected and Pneumonia. The deep learning used the DarkNet neural network. The data was divided between 80 \% training and 20 \% testing. The deep learning was ran for 100 epochs using 5 fold cross validation. The results produced an accuracy score of 98 \% for binary classification and an accuracy score of 87.02 \% for multi class classification. It is expected that the result will fall as the number of classifications increase as the AI will need to recognise more features to distinguish between classes. The complete process followed in this work is detailed in figure \ref{ozturk2020automated1}.

\subsection{Camera Technology}
Camera technology can be used to provide non contact sensing by observing the chest movements of an individual \cite{nam2016monitoring}. This can provide assistance in the detection of COVID-19 as one of the symptoms include an increase in breathing rate. 

The paper "Combining Visible Light and Infrared Imaging for Efficient Detection of Respiratory Infections such as COVID-19 on Portable Device" used RGB-thermal camera footage for the detection of COVID-19. The footage was used with machine learning to detect normal and abnormal breathing from people wearing protective masks. The research collected real world data and applied deep learning to achieve a good result of 83.7 \% accuracy which is the highest result in the literature in regards to breathing detection using RGB-thermal cameras using deep learning. This research can provide a scanning method which can be used to control the spread of the virus \cite{jiang2020combining}. 

\begin{figure}[htb]
    \centering
    \includegraphics[width=0.5\textwidth]{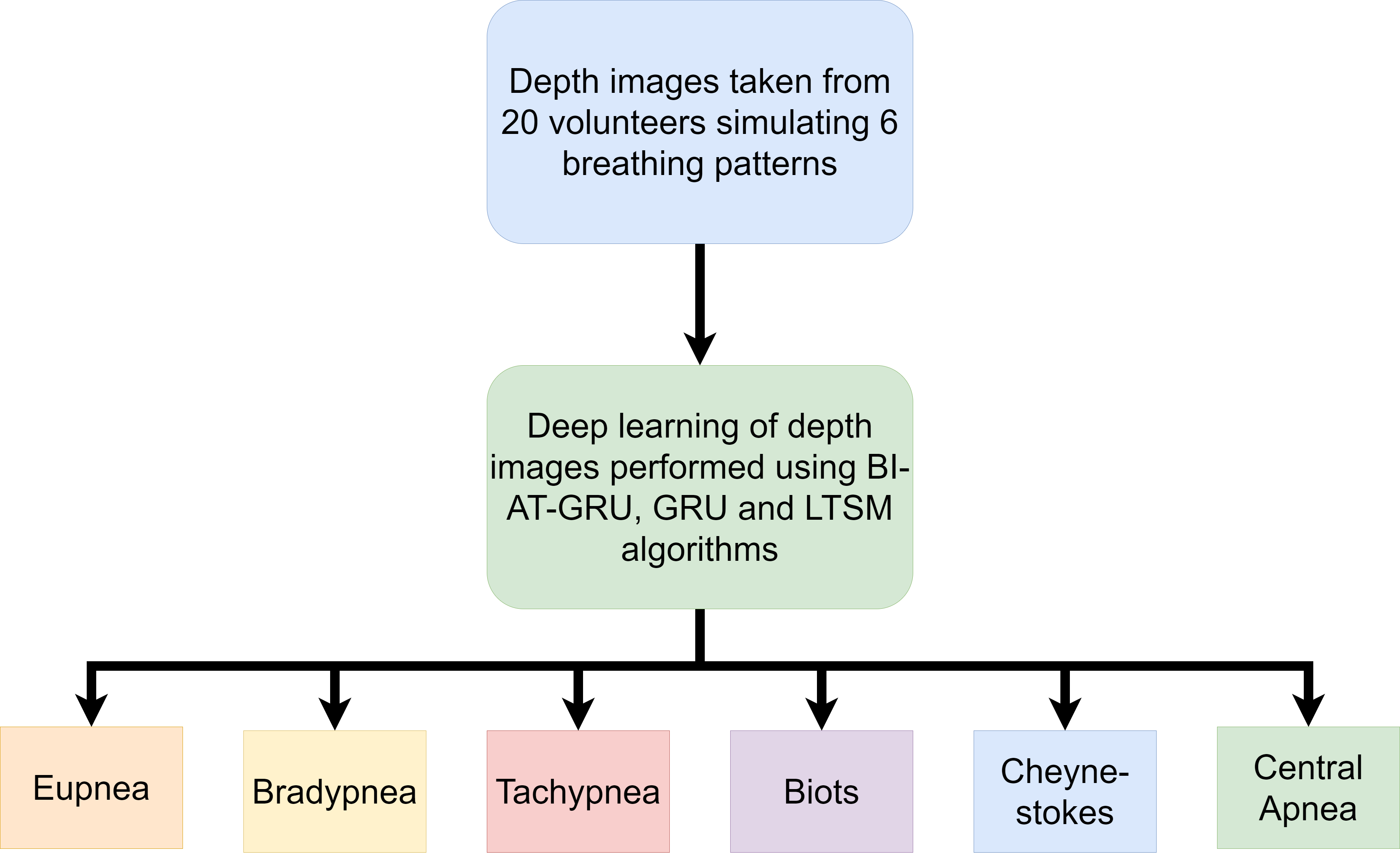}
    \caption{Flow chart of work for detection of COVID-19 from Depth Camera Image (Reproduced from \cite{wang2020abnormal})}
    \label{wang2020abnormal1}
\end{figure}

Wang, Yunlu, et al. \cite{wang2020abnormal} used Microsoft Kinect cameras to take depth images of volunteers breathing. A total of 20 volunteers were asked to sit on a chair and simulate 6 different breathing patterns. The breathing patterns were Eupnea, Bradypnea, Tachypnea, Biots, Cheyne-stokes and Central Apnea. Patients of COVID-19 display the rapid breathing pattern of Tachypnea. During data collection, a Spirometer was used to ensure the breathing pattern was being simulated correctly. The images taken using the camera are used in a deep learning neural network to classify abnormal breathing patterns associated with COVID-19. The deep learning model used was the BI-AT-GRU algortithm. Gated Recurrent Unit (GRU) is a simplified version of Long Term Short Memory(LTSM) algorithm. The BI-AT-GRU algorithm results achieved the best accuracy score of 94.5 \%. This research shows how depth images can be used to identifiy the Tachypnea breathing pattern observed in COVID-19 patients. The process map for this research is shown in figure \ref{wang2020abnormal1}.

The primary disadvantage of using this method is the cost of thermal and depth cameras and the camera operators. Although the price of these cameras is falling gradually, it still remains substantially high \cite{elphick2019exploratory}. The reseach done with cameras has shown that the devices can be used with AI in the detection of COVID-19 without contact with the body.

\subsection{Ultrasound Technology}
 Ultrasound technology can be applied to detect respiratory failure of the lungs. An ultrasound machine is a device that uses high frequency sound waves to image body movements \cite{powles2018physics}. The sound waves bounce off different parts of the body which create echoes that are detected by the probe and used to create a moving image. Lung ultrasounds has seen great development in  recent years \cite{mojoli2019lung}. The use of Ultrasound Technology can be used in the detection of COVID-19 in a non contact method where the risk of healthcare professionals becoming infected from patients can be decreased \cite{soldati2020there, buonsenso2020covid}. Ultrasound technology can be performed using smart phones for the signal and processing of ultrasound images in a portable setting \cite{kim2013smartphone}. The disadvantage of ultrasound technology is that patients must prepare themselves before an ultrasound can effectively create an image of the body \cite{genc2016ultrasound}. This preparation can include not eating for a few hours before.

\begin{figure}[htb]
    \centering
    \includegraphics[width=0.5\textwidth]{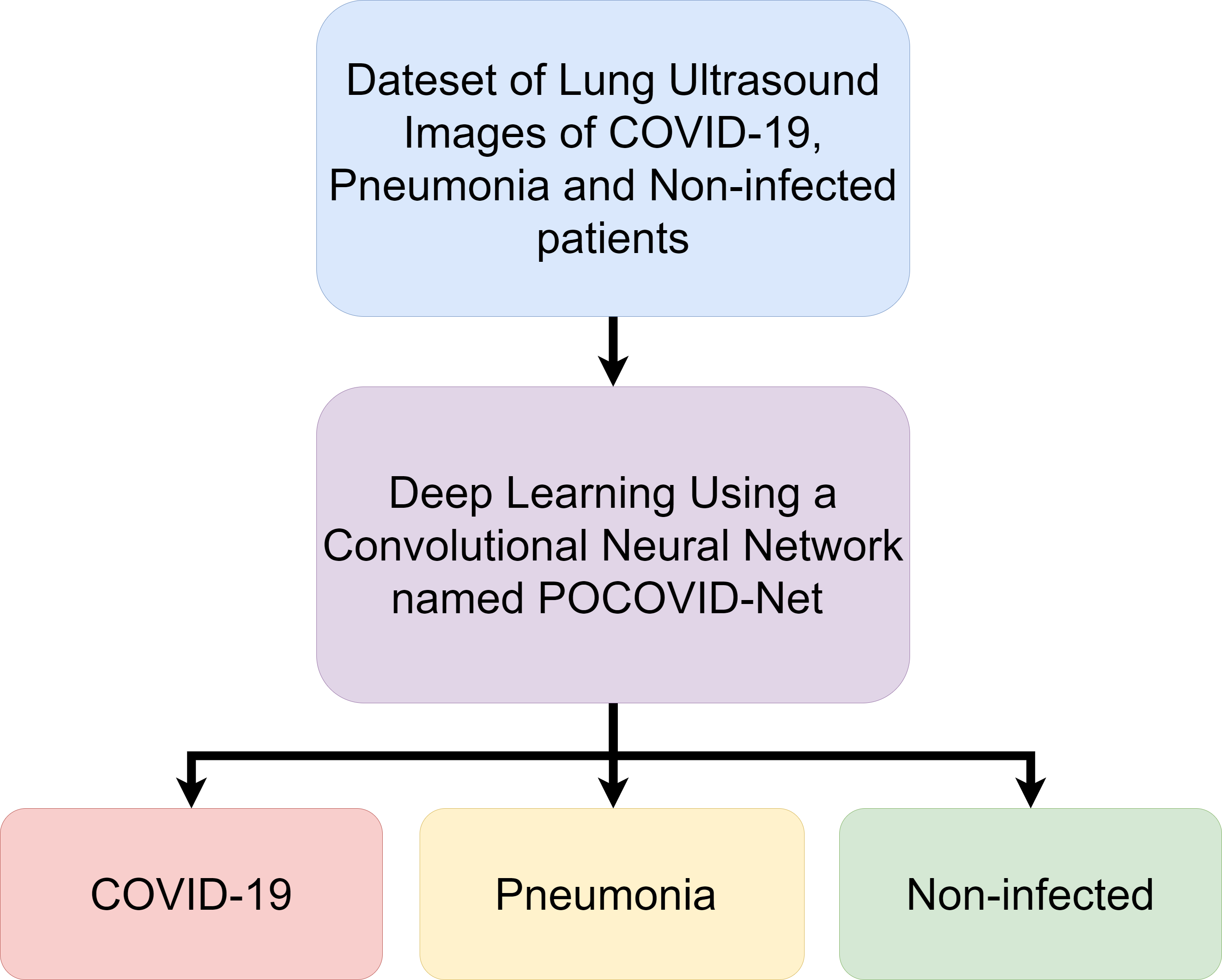}
    \caption{Flow chart of work for detection of COVID-19 from Ultrasound Technology (Reproduced from \cite{born2020pocovid})}
    \label{born2020pocovid1}
\end{figure}

The work of Born, Jannis, et al. \cite{born2020pocovid} shows that ultrasound technology can be used with deep learning to distinguish the differences in COVID-19, Pneumonia and no infection. The research collects a dataset of lung ultrasound images which contain video recordings of lung ultrasound scans. The dataset includes a total of 64 video recordings with 39 of the recordings of COVID-19 patients, 14 videos of Pneumonia patients and 11 videos of non-infected patients. The paper has developed a convolutional neural network named POCOVID-Net. The deep learning algorithm was able to achieve an accuracy score of 89 \%. Figure \ref{born2020pocovid1} shows a simplified flow graph of the experiment undertaken in this paper. 

\subsection{Radar Technology}
Radar Technology can be used to monitor the respiratory system within a home environment and provide a quick response if abnormalities are found which suggest COVID-19 being present. Radar systems use frequency-modulated continuous-wave (FMCW) to observe the Doppler effect when a person moves \cite{ding2019fall,shah2019cognitive,yang2020diagnosis,shah2020sensor}. This can be used to monitor the fine movements associated with breathing. Research done shows that radar technology can achieve 94 \% accuracy for the detection of breathing rates and 80 \% accuracy for heart rate detection \cite{alizadeh2019remote,fioranelli2019radar,fan2018breathing}. The Israeli military force has made use of radar systems for monitoring the vital signs of COVID-19 patients. The goal of using this method is to prevent medical staff from becoming infected while caring for patients \cite{islamcan,shah2019human}. Using radar technology to monitor vital signs can provide non-interference monitoring, however the disadvantages of radar systems is that it has high power requirements and the technology comes at a high cost \cite{christenson2019low}. 

\subsection{Radio Frequency Signals}
The use of radio frequency (RF) signal sensing can be used to detect the vital signs of individuals by sensing the minute movements of the chest made while breathing as the heart beats ( \cite{liu2018monitoring,dong2017monitoring,haider2018utilizing,yang2018detection,yang2018monitoring}). This technique can be used for monitoring the vital signs of patients independent of their activities \cite{liu2015tracking}. The RF signals detect the movement by observing the Channel State Information (CSI) which can show amplitudes of the RF signals while movement occurs between a RF transmitter and receiver. \cite{zhao2019r,chopra2016thz}. The Emerald system has been developed  to monitor patients using RF signals. The system uses RF signals to detect the breathing rate of COVID-19 patients and then uses AI to infer the breathing rate of the patient. This allows for doctors treating the patients to be able to monitor the patient from a safe distance. This method prevents the risk of infection to staff and provides the patient comfort as they do not need to wear monitoring devices \cite{conner-simons_2020}. RF signals can be vulnerable to other movements within the room. The other movements create noise in the CSI which can then in turn cause false readings.

\subsection{Future Directions}
This section will detail some of the future directions which may be suitable for expanding on the research presented in this paper. The research has highlighted how the detection of COVID-19 is possible using various techniques. This section will now discuss how this research can be taken further to work within real life scenarios.

\begin{itemize}

\item One of the biggest challenges with CT scanning as a means to diagnose COVID-19 is the lack of portability. This means that although the method is non contact, its use still requires individuals to travel to a location where the machine is available. As the CT images are able to provide high resolution, the AI can be used to for the detection of COVID-19. Therefore future directions of this method should look to creating highly accurate models that can eventually lead to the automation of COVID-19 detection. This can allow for faster diagnosis. Which can allow for more patients to be tested and increase availability of staff operating and analysing CT scans.

\item X-rays similarly to CT scans are not portable. Like CT scans professionals are required to operate these machines and analyse the X-ray images. The research presented in this paper has shown that AI can be used to make predictions if COVID-19 is present in the lungs. This can be useful similarly to CT scans where AI can be applied to make the predictions and speed up the process. The more data collected, the more advanced the model will become. Perhaps initially the predictions will need to be confirmed by humans but eventually the checks can become less frequent. Since the research above have displayed an ability of AI to distinguish between not just COVID-19 and non infected but also pneumonia at high accuracy, then the AI has proved to be capable of accurate classifications.

\item Thermal and Depth cameras are able to detect the irregular breathing patterns that are associated with COVID-19 symptoms. The issue here is that even though the camera can detect the irregular breathing pattern it is unable to categorically define COVID-19 as the cause for individuals displaying the irregular breathing patterns. In a real life situation, the camera method may be better suited to monitoring vulnerable people who are considered high risk from COVID-19. Then once the monitoring system has identified the irregular breathing patterns and alarm can be raised with a career or family member. Then appropriate action can be taken for greater accuracy such as diagnosis with CT scanning or X-ray scanning.

\item Ultrasound technology is able to take moving images of the lungs and detect COVID-19. This can also be made portable by using mobile devices. AI can be applied to recognise if COVID-19 or pneumonia is present in the lungs. This research can be further applied to develop applications on a mobile that device can capture an ultrasound of the lungs then compared to a AI model to predict if COVID-19 is present. Although not all phones may not have the necessary hardware to achieve this, the non contact method can allow for others to be able to use the devices for diagnosis at a safe distance. 

\item Radar technology is able to identify the breathing patterns of individuals. Much like camera technology, the identification of breathing patterns are able to raise cause of concern but it cannot isolate COVID-19 as the sole cause. Radar technology can again be used to monitor individuals but due to the high costs its more likely to be used as a monitoring system within a hospital and not a home environment. 

\item Any future directions should consider the use of RF signals as a means to detect the breathing patterns which give indication of COVID-19 symptoms. The RF systems can be implemented inexpensively using existing WiFi technology present within many homes. This allows for monitoring of individuals without the costs incurred in implementing radar or camera technologies highlighted in this paper. 

\end{itemize}

\section{Conclusion}
The works listed in this paper have shown that the COVID-19 virus can be detected using contact-less techniques. Techniques such as CT scans and X-ray imaging provide high accuracy and high image resolution but the cost of the equipment is high and not portable. Thermal and depth camera technology has been used to detect breathing patterns which is associated with COVID-19 symptoms. However these cameras are expensive and needs to be operated by a professional. Radar technology also able to detect the breathing patterns but carries disadvantages of high operating expenses and capital expenditures. RF signals provide low cost and high accuracy as compared with other non-invasive technologies. The technologies are able to work on AI which can allow for skilled professionals to be available to assist in other areas of healthcare during the pandemic. The non-contact methods also protect healthcare workers from contracting the disease. The future direction of non-contact detection should look at the use of RF systems as the cost is cheap and it is easier to implement within a home environment in comparison to other methods. This gives the advantage of allowing the users to remain within isolation.

\section{Acknowledgement}
William Taylor’s studentship is funded by CENSIS UK through Scottish funding council in collaboration with British Telecom. This work is supported in parts by EPSRC DTG EP/N509668/1 Eng, EP/T021020/1 and EP/T021063/1

\printbibliography

\end{document}